\documentclass{PoS}

\usepackage{graphicx,amsmath,amssymb,cite}

\usepackage{psfrag}
\usepackage{epsfig}
\newcommand{\be}{\begin{eqnarray}}
\newcommand{\ee}{\end{eqnarray}}
\newcommand{\beq}{\begin{equation}}
\newcommand{\eeq}{\end{equation}}

\newcommand{\mpi}{M_{\pi}}

\title{Pion reactions with few-nucleon systems}

\ShortTitle{Pion reactions with few-nucleon systems}

\author{\speaker{Vadim Baru},\\
        Forschungszentrum J\"ulich, Institut f\"ur Kernphysik (Theorie) and  J\"ulich Center for
Hadron Physics,  D-52425 J\"ulich, Germany and\\
          Institute for Theoretical and Experimental Physics, B. Cheremushkinskaya 25,\\ 117218 Moscow, Russia\\
                E-mail: \email{v.baru@fz-juelich.de}}

\abstract{
We  report about the recent results for s- and p-wave pion production in 
$NN\to NN \pi$ within effective field theory and 
discuss how the charge symmetry breaking in $pn\to d\pi^0$ can be used to extract the strong contribution to 
the neutron-proton mass difference.
}

\FullConference{6th International Workshop on Chiral Dynamics, CD09\\
		July 6-10, 2009\\
		Bern, Switzerland}

\begin{document}

\vspace*{-1cm}
\section{Introduction}
\label{intro}
With the advent of chiral perturbation theory (ChPT), the low-energy effective
field theory (EFT) of QCD, high accuracy calculations for
hadronic reactions with a controlled error estimation have become
possible~\cite{chpt1,chpt2}.
The framework  has been successfully applied to study, in particular, $\pi\pi$~\cite{pipi} and 
$\pi N$~\cite{piN} scattering observables as well as  nuclear forces~\cite{NN}.
In this contribution we discuss an application of ChPT to 
the reactions involving pion production 
on two-nucleon systems.  This type of reactions  
 allows one to test predictions of ChPT in the process with the large momentum 
transfer typical for the production process. 
As was first advocated in Refs.~\cite{bira1,rocha}, 
the  initial nucleon momentum in the threshold kinematics 
sets the new "small" scale in the problem, namely,  
$p\simeq \sqrt{m_N \, M_\pi} \simeq 360\, $MeV  ($\chi\simeq p/m_N\simeq \sqrt{\mpi/m_N}$), 
where $\mpi$ ($m_N$) is the pion (nucleon) mass.
The proper way to include this scale in the 
power counting was presented in Ref.~\cite{ch3body} and implemented in
Ref.~\cite{withnorbert}, see Ref.~\cite{report} for a review
article.  As a consequence, the hierarchy of diagrams changes 
in the modified power counting scheme of Ref.~\cite{ch3body},
and some loops start to contribute already at NLO
for $s$-wave pion production. On the other hand,  
$p$-wave pion production is governed by the 
tree-level diagrams up to NNLO. In what follows we discuss the status of the theory for 
pion production in the isospin conserving and isospin violating case. 
It was first argued in Ref.~\cite{kolck} that charge symmetry breaking (CSB) effects 
recently observed experimentally in $pn\to d\pi^0$  \cite{Opper}
provide an access to the neutron-proton  
mass difference which is the fundamental quantity of QCD.
To take this challenge, however, one needs to have the isospin conserving case fully under control.
In sec.2 we briefly discuss the theoretical status  for s-wave pion production. 
Sec. 3 is devoted to a more detailed discussion of the recent results for  p-wave pion production.
In sec. 4 we  briefly highlight the recent developments in the study of charge symmetry 
breaking effects in $pn\to d\pi^0$. We finalize with the summary of the latest results.

\section{s-wave pion production and the concept of reducibility}
\label{s-wave}
\begin{figure}[h!]
\begin{center}
\includegraphics[height=5.5cm,width=14.0cm,keepaspectratio]{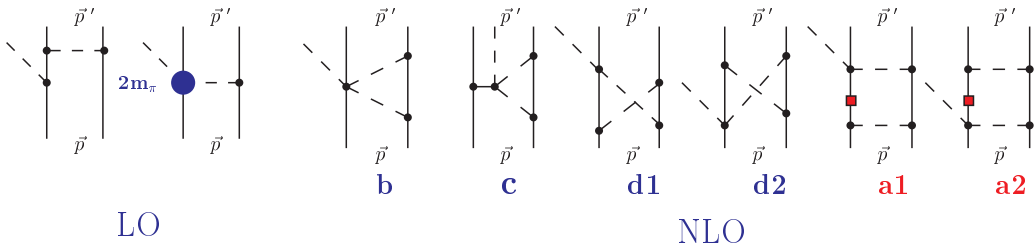}
\caption{Complete set of nucleonic diagrams up to NLO. 
Note that sum of all loops at NLO vanishes.}
\label{NLOdiagr}
\end{center}
\end{figure}
A method  how to calculate processes on few
nucleon systems with external probes was proposed by Weinberg \cite{swein1}: 
\begin{enumerate}
\item the perturbative transition (production) operators have to be calculated systematically
using ChPT. They should consist of irreducible graphs only. 
\item the transition operators have to be convoluted with the non-perturbative $NN$ wave
functions.
\end{enumerate}
Therefore it is necessary to disentangle those diagrams that are part of the
wave function from those that are part of the transition operator. In complete
analogy to $NN$ scattering, the former are called reducible
and the latter irreducible. The distinction stems from whether 
the diagram shows a two-nucleon cut or not.
Thus, in accordance to this rule,
 the one-loop diagrams shown in
Fig. \ref{NLOdiagr}(b)--(d) are irreducible, whereas diagrams (a)  seem to be
reducible. This logic was used in Ref.~ \cite{withnorbert}  
to  single out the irreducible loops contributing at NLO.
The findings of Ref.~\cite{withnorbert} were:
\begin{itemize}
\item For the channel $pp\to pp\pi^0$ the sum of diagrams (b)--(d) of Fig.
  \ref{NLOdiagr} vanished due to a cancellation between individual diagrams
\item For the channel $pp\to d\pi^+$ the same sum gave a finite answer\footnote{The
connection of the amplitude $A$ to the observables is given, e.g., in Ref~\cite{lensky2}}:
\begin{equation}
A_{pp\to d\pi^+}^{b+c+d} = \frac{g_A^3}{256f_\pi^5} \left(-2+3+0\right)\,
|\vec q|=\frac{g_A^3|\vec q|}{256f_\pi^5},
\end{equation}
where $f_\pi$ denotes the pion decay constant and 
 $g_A$ is the axial-vector coupling of the nucleon.
\end{itemize}\noindent
\begin{figure}[t!]
\parbox{0.5\textwidth}{\hspace*{-0.2cm}\includegraphics[scale=0.5,clip=]{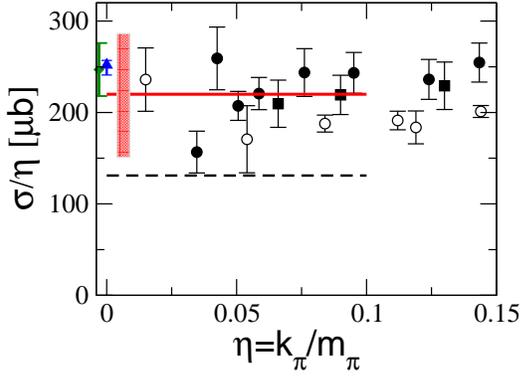}}
\parbox{0.5\textwidth}{\vspace*{-1.5cm}\caption{Comparison of our results
 to experimental data for $NN\to d\pi$. The data sets are from Refs.~\cite{dpidata1,dpidata2,dpidata3,dgotta,Strauch}.
Note that the green diamond and the blue triangle correspond to the most resent measurements from pionic deuterium atom \cite{dgotta,Strauch}
 at $\eta=0$.  The dashed line corresponds to the model of Koltun and Reitan~\cite{KR}. The solid red curve represents 
our results, as given in Ref.\cite{lensky2}, the filled red box demonstrates the theoretical uncertainty of the NLO calculation. 
}}
\label{NNpi_result}
\end{figure}
The latter amplitude grows linearly with increasing final
 $NN$--relative momentum  $|\vec q|$, which leads to a large
  sensitivity to the final $NN$ wave function, once the convolution of those
  with the transition operators is evaluated. However, the problem is that such a
 sensitivity is not allowed in a consistent field theory as was stated in Ref.~\cite{Gaa05}. 
The solution of this problem was presented in Ref.~\cite{lensky2} (see also Ref.\cite{NNpiMENU}). 
It was pointed out in Ref.~\cite{lensky2} that the diagrams that look formally reducible 
can acquire irreducible contributions in the presence of the energy-dependent  vertices 
(or time-dependent Lagrangian densities). 
Specifically, the energy dependent part of the leading Weinberg-Tomozawa (WT) $\pi N\to \pi N$ 
vertex cancels one of the 
intermediate nucleon propagators (see the one with the red square in Fig.~\ref{NLOdiagr}), 
resulting in an additional irreducible 
contribution at NLO. 
It turned out that this additional irreducible contribution compensates the linear 
growth of diagrams (b)--(d) thus solving the problem. 
Thus, up to NLO, only the diagrams appearing at LO (see Fig. \ref{NLOdiagr}), 
contribute to $pp\to d\pi^+$, with the rule that the $\pi N\to \pi N$ vertex is put 
on--shell. The latter stems from the observation that in addition to the leading WT-term ($\sim 3\mpi/2$) 
the nucleon recoil  correction to the WT $\pi N$ vertex
($\sim \mpi/2$) also contributes at LO. 
As a result  the dominant $\pi N$-rescattering amplitude is enhanced by a 
factor of $4/3$ as compared to the traditionally used value ($\sim 3\mpi/2$), 
which leads to a good description of the experimental data for 
$pp\to d\pi^+$ (see Fig.~\ref{NNpi_result}). 
Note, however, that the relatively large theoretical uncertainty of about $2\mpi/m_N\approx 30\%$ 
for the cross section requires a carefull study of higher order effects.

\vspace*{-0.2cm}
\section{p-wave pion production}
\label{p-wave}
\begin{figure}
\centerline{\epsfig{file=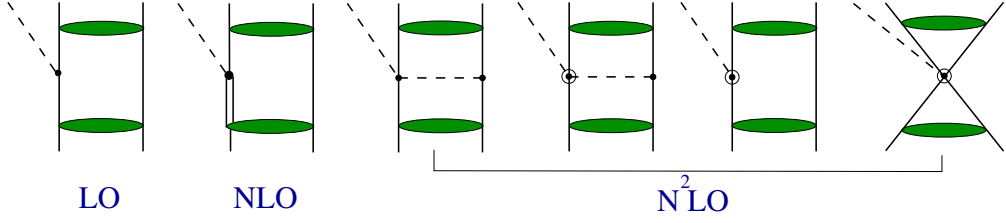, height=2.8cm}}
\caption{Diagrams that contribute  to
the $p$-wave amplitudes of $NN\to NN\pi$ up to NNLO.}
\label{diag_pwave}      
\end{figure}
Diagrams that contribute to p-wave pion production up to NNLO in the modified power counting are shown in Fig.\ref{diag_pwave}.
In particular, at NNLO there are subleading rescattering and direct pion production operators as well as 
the $(\bar NN)^2\pi$ contact term. Notice that it is the same contact term that also contributes  to 
the three-nucleon force  \cite{ch3body,pdchiral},  to the processes $\gamma d\to\pi
NN$~\cite{gammad,gammad_nn} and $\pi d\to \gamma NN$~\cite{gardestig,gardestig2}
as well as to weak reactions such as, e.g., tritium beta
decay and proton-proton $(pp)$ fusion~\cite{park2,phillips}. 
Therefore it provides an important connection between different low-energy reactions. 
It is getting even more intriguing once one realizes that this
operator appears in the above reactions in very different
kinematics, ranging from very low energies for both incoming and
outgoing $NN$ pairs in $pd$ scattering and the weak reactions up to
relatively high initial energies for the $NN$ induced pion production.
As a part of this connection in Ref.~\cite{gazit} it was shown that both the $^3$H and $^3$He binding energies
and the triton $\beta$-decay can be described with the same contact term. 
However, an apparent discrepancy between the strength
of the contact term needed in $pp\to pn\pi^+$ and in $pp\to de^+ \nu_e$ 
was reported in Ref.~\cite{nakamura}.
\begin{figure}[t]
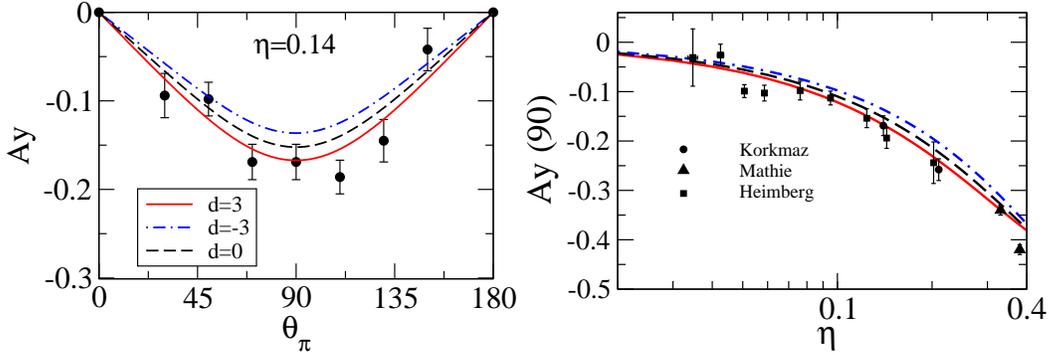

\begin{center}
\vspace*{.2cm}
\includegraphics[width=0.45\textwidth,clip=]{Ay_0_14.eps}
\includegraphics[width=0.46\textwidth,clip=]{Ay90dpi_with_pf_fully_eta04v0.eps}
\caption{Results for the analyzing power at $\eta$=0.14 (left panel)
 and  the analyzing power at 90 degrees
(right panel) for the reaction $pp\to d\pi^+$ for different
values of the LEC $d$  (in units  $1/(f^2_{\pi}M_N)$)
of the $(\bar NN)^2\pi$ contact operator. Shown are $d=3$ (red solid line), 
$d=0$ (black dashed line), and $d=-3$ (blue dot-dashed line). 
The data are from Refs.~\cite{Ritchie,Heimberg,Drochner,Korkmaz,Mathie}.
}
\label{Ay_dpi}
\end{center}
\end{figure}
 If the latter observation were true, it would 
 question the applicability of chiral EFT to the reactions $NN\to NN\pi$.
To better understand the discrepancy reported in Ref.~\cite{nakamura}, in the recent 
paper \cite{we_pwave} we simultaneously analyzed 
different pion production channels. In particular, we calculated the
$p$-wave amplitudes for the reactions $pn\to pp\pi^-$, $pp\to
pn\pi^+$, and $pp\to d\pi^+$. Note that even in these channels the contact
term occurs in entirely different dynamical regimes. For
the first channel $p$-wave pion is produced  along with the
slowly moving protons in the ${^1\!}S_0$ final state whereas for the other
two channels the ${^1\!}S_0$ $pp$ state is to be evaluated at the
relatively large initial momentum. 
In practice, 
 we varied  the value of the low-energy constant (LEC) $d$, which represents the strength of the 
contact operator,  in such a way to get the 
best simultaneous qualitative description of all channels of $NN\to NN\pi$. It should be stressed, however, that 
the value of $d$ depends on the $NN$ interaction employed and on the method used to regularize the overlap
integrals.  It therefore does not make much sense to compare values for $d$ as found in
different calculations. 	Instead one should  compare results on the level of observables and
this is what we  do below (see also Ref.~\cite{we_pwave}).
In Fig.\ref{Ay_dpi} we compare our results for various values of $d$ with the experimental data 
for the analyzing power for the reaction $pp\to d\pi^+$. We find that the data prefer a positive value of 
$d$ of about 3. A similar pattern can be observed in the reaction $pn\to pp\pi^-$ as illustrated in Fig.~\ref{res_pppimin}.
Again the data show a clear preference of the positive value for LEC $d$ -- our fit resulted in $d=3$ for the 
best value.  This channel, however, has been measured at TRIUMF 
at relatively large excess energy ($\eta= 0.66$) where the conclusion may 
be spoiled due to the onset of  pion d-waves. 
Fortunately, a new  measurement for the same observables at lower energies is currently ongoing
at COSY~\cite{ANKE} which  will soon allow a  quantitative extraction
of the value of the LEC $d$.
\begin{figure}[t]
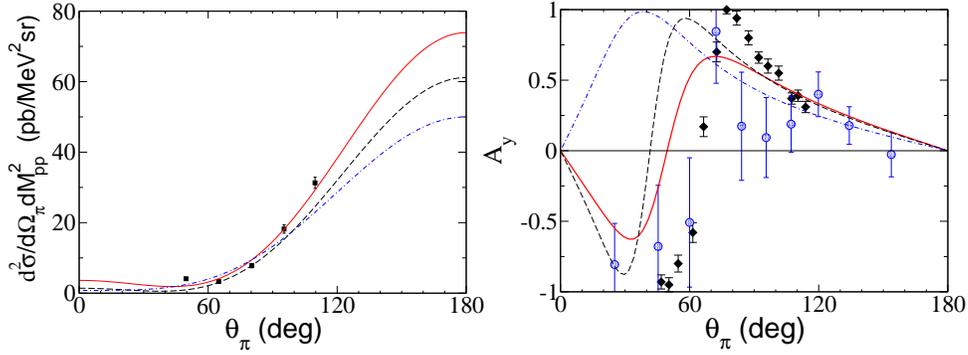

\begin{center}
\includegraphics[width=0.42\textwidth,clip=]{Fig6a.eps}\hspace*{-0.12cm}
\includegraphics[width=0.425\textwidth,clip=]{Fig6b.eps}
\caption{ 
Results for $d^2\sigma/d\Omega_{\pi}dM^2_{pp}$ (left panel) and $A_y$
(right panel) for $pn\to pp(^1S_0)\pi^-$. 
Shown are the results for $d=3$ (red solid line), $d=0$ (black dashed line) and $d=-3$
(blue dot-dashed line). 
The data are from  TRIUMF~\cite{hahn,duncan} (black squares) and from
PSI~\cite{Daum} (blue circles) .} 
\label{res_pppimin}  
\end{center}
\end{figure}
We now turn to the reaction $pp\to pn\pi^+$  -- this channel was used  in the analysis of Ref.\cite{nakamura}.
The reaction $pp\to pn\pi^+$ should have, in principle, the same information on the LEC $d$
 as contained in the  deuteron channel. However, it is much more difficult to extract the pertinent 
information unambiguously from this reaction. 
In particular, the final NN-system might 
be not only in $S$- but also in $P$-wave both for isospin-zero and for  isospin-one NN states. 
At the energies considered in the experimental investigation, $\eta=$0.22, 0.42, and 0.5, the $Pp$
amplitudes may contribute significantly~\cite{bilger,complete,pwpi0}. In the current study these states 
are disregarded. The results of our calculation for the magnitude $A_2$
 are given\footnote{
The coefficients $A_i$ are related to the unpolarized differential cross section via
$\frac{d\sigma}{d\Omega} = A_0 + A_2 P_2(x) \ ,$
with $P_2(x)$ being the second Legendre polynomial }
 in the left panel of Fig.~6 . One finds again that the positive LEC $d\simeq 3$ seems to be preferred.
 Thus we conclude that all reaction channels of $NN\to NN\pi$ can be described simultaneously with the same value of the 
 LEC $d$. Coming back to the problem raised in Ref.~\cite{nakamura} it should be pointed out that the results of this work were
 not compared directly to the observables in $pp\to pn\pi^+$. Instead, they were compared to the results of the partial wave analysis (PWA)
 performed in Ref.\cite{Flammang}, as demonstrated in the right panel of Fig.~6. It is based on this discrepancy between 
 data and theory it was concluded in Ref.\cite{nakamura} about the failure of simultaneous description of the weak processes and $NN\to NN\pi$.
 However, the partial wave analysis of Ref.\cite{Flammang} seems to be  not correct. Here we refer the interested 
 reader to Ref.\cite{we_pwave} where  the drawbacks of this PWA are discussed in detail. 
To illustrate the problems of the PWA in the right panel of 
Fig.~6 we also show the results of our calculation for the relevant partial wave $a_0$ which corresponds to the  transition
$^1S_0\to {}^3S_1p$ where the contact term acts. 
Clearly, although all data presented in Ref.~\cite{Flammang} 
are in a good agreement with our calculation (see left panel in
Fig.~6 and also Ref.\cite{we_pwave} for more details), 
the partial wave amplitude $a_0$ is not at all described. Thus, we think 
that the problem with the simultaneous description of  $pp\to de^+ \nu_e$ 
 and $pp\to pn\pi^+$, raised  in Ref.\cite{nakamura},
 is  due to the drawbacks of the partial 
wave analysis of Ref.\cite{Flammang}. 
\begin{figure}[tb]
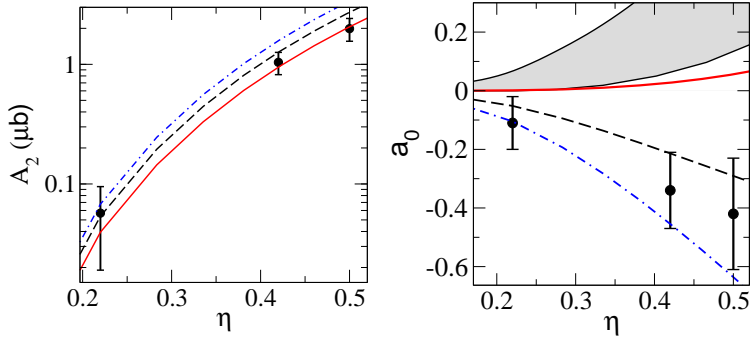
\hspace*{-0.3cm}
\parbox{0.33\textwidth}{\hspace*{-0.2cm}\includegraphics[scale=0.39,clip=]{Fig7a.eps}}
\parbox{0.33\textwidth}{\hspace*{-0.2cm}\includegraphics[scale=0.37,clip=]{a0ct_with_pf_paperv2.eps}}
\parbox{0.33\textwidth}{\caption{ 
Results for the magnitude $A_2$ (left panel)
and the  partial wave amplitude  $a_0 (\sqrt{\mu b})$ representing the relevant transition 
$^1S_0\to {}^3S_1p$  (right panel) for 
$pp\to pn\pi^+$ for different values of the LEC $d$. 
The notation is the same as in Fig.~5, gray band corresponds to the results of Ref.~\cite{nakamura}. 
 The data are from Ref.~\cite{Flammang}.
}}\label{A2}
\end{figure}

\vspace*{-0.5cm}
\section{CSB effects in $pn\to d \pi^0$}

Recently, experimental evidence for CSB was found in reactions
involving the production of neutral pions. At IUCF non-zero values for the
$dd\to \alpha \pi^0$ cross section were established \cite{Stephenson}. 
At TRIUMF a forward-backward
asymmetry of the differential cross section for $pn\to d\pi^0$
was reported which amounts to $A_{fb}=[17.2
\pm 8 {\rm (stat.)} \pm 5.5 {\rm (sys.)}] \times 10^{-4}$~\cite{Opper}.
In a charge symmetric world the initial $pn$ pair would 
consist of identical nucleons in a pure isospin one state. 
Thus the apparent forward--backward asymmetry is due to charge symmetry breaking.

\begin{figure}[tb]
\parbox{5.5cm}{\hspace{0cm}\includegraphics[scale=0.65,clip=]{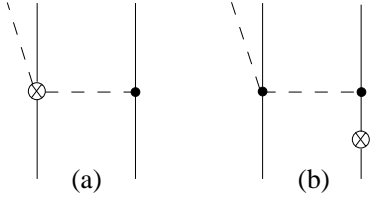}}
\parbox{9cm}{\hspace*{-0.5cm}
\caption{Leading order diagrams for the isospin
  violating $s$-wave amplitudes of $pn\to d\pi^0$.
Diagram (a) corresponds to isospin
 violation in the $\pi N$ scattering vertex explicitly whereas  diagram (b)
 indicates an isospin-violating contribution due to the neutron--proton 
 mass difference in conjunction with the time-dependent
  Weinberg-Tomozawa operator.  }}
\label{diagLO}
\end{figure}

The neutron--proton mass difference is due to strong and electromagnetic
interactions \cite{Gasser}, i.e. 
$\delta m_N=m_n-m_p=\delta m_N^{\rm str}+\delta m_N^{\rm em}$. It was stressed and exploited 
in Ref.~\cite{kolck} that  the
strength of the  rescattering CSB operator at LO  in $pn\to d\pi^0$ (see  Fig.~\ref{diagLO}(a))
is proportional to a different combination of $\delta m_N^{\rm str}$ and 
$\delta m_N^{\rm em}$  ( see also \cite{ulfsven,Fettes:2000vm} for related works). 
Thus the analysis of CSB effects in $pn\to d\pi^0$ should allow
to determine these important quantities 
individually.
It was, however, quite surprising to find that, using
the values for $\delta m_N^{\rm str}$ and $\delta
m_N^{\rm em}$ from Ref.~\cite{Gasser}, the leading order calculation
of the forward-backward asymmetry \cite{kolck} over-predicted the
experimental value by about a factor of 3 --- a consistent description
would call for an agreement with data within the theoretical
uncertainty of 15\% for this kind of calculation. The evaluation of certain
higher order corrections performed in Ref.~\cite{kolck} and in a 
recent study~\cite{bolton} did not change the situation sufficiently
--- the significant overestimation of the data persisted.
In the recent work \cite{we_CSB} 
 we have shown that there is one more rescattering
operator that contributes at LO (see  diagram (b) in Fig.~\ref{diagLO}). 
In full analogy to isospin conserving s-wave pion production, 
the idea was based  on the fact that the  energy-dependent WT $\pi N$ vertex acquires 
an additional contribution proportional to $\delta m_N$ as soon as we distinguish between the
proton and the neutron. We evaluated this new LO operator 
and  recalculated the forward-backward assymetry at LO. It should be pointed out at this stage
that $A_{fb}$ at LO  is proportional  to the interference of the s-wave pion CSB amplitude at LO and 
the p-wave pion isospin conserving amplitude. The latter is calculated 
up to NNLO, as discussed in Sec.\ref{p-wave}, and exhibits very good description of data, which is 
a necessary pre-requisite for studying CSB effects. 
The complete LO calculation gives \cite{we_CSB}
\beq 
A_{\rm fb}^{\rm LO} = (11.5 \pm 3.5)\times 10^{-4} \ 
\frac{\delta m_N^{\rm str}}{\rm MeV} \ 
\label{AfbLO}
\eeq
which agrees nicely  with the experimental
data if one uses the value of $\delta m_N^{\rm str}$ from Ref.\cite{Gasser}.  
We may also use (Eq.~\ref{AfbLO})  to extract  $\delta m_N^{\rm str}$ 
from the above expression using the data, which yields 
\beq
\delta m_N^{\rm str} = 
\left( 1.5 \pm 0.8 \ {\rm (exp.)} \pm 0.5 \ {\rm (th.)}
\right) \ {\rm MeV}.
\eeq
This  result reveals a  very good agreement with the one based on the Cottingham sum rule \cite{Gasser},
$\delta m_N^{\rm str} = 2.0 \pm 0.3  \ {\rm MeV}\,$,
and with a recent determination of the same quantity using lattice QCD~\cite{lattice},
$\delta m_N^{\rm str}=2.26 \pm 0.57 \pm 0.42 \pm 0.10$ MeV.

\vspace*{-0.5cm}
\section{Summary}
\label{summary}
We  surveyed the recent developments  for 
$NN\to NN \pi$. 
We showed, in particular, (see Ref.\cite{lensky2,NNpiMENU}) that 
the s-wave pion production amplitudes calculated up to NLO for $pp\to d\pi^+$
provide a good qualitative understanding of the pion dynamics. However, 
the relatively large theoretical uncertainty of about $2\mpi/m_N\approx 30\%$ 
for the cross section requires a computation of loops at NNLO.   The latter are also absolutely 
necessary for $pp\to pp\pi^0$, see Ref.~\cite{myhrer} for the first results in this direction. 
Recently, we have studied $p$-wave pion production  up to NNLO \cite{we_pwave}. In particular, we showed that
it is possible to describe simultaneously the $p$-wave amplitudes
in the $pn\to pp\pi^-$, $pp\to pn\pi^+$, $pp\to d\pi^+$ channels by 
adjusting a single low-energy constant accompanying the short-range $(\bar NN)^2\pi$ operator
 available at NNLO. We also demonstrated that the problem with the simultaneous description of 
the weak proton-proton fusion process and $pp\to pn\pi^+$, reported in Ref.\cite{nakamura},
 is most probably due to the drawbacks of the partial 
wave analysis of Ref.\cite{Flammang} used in Ref.\cite{nakamura}. 
Based on good  understanding of the pion production mechanisms in the isospin 
conserving case we studied charge symmetry breaking effects in $pn\to d\pi^0$. We performed 
a complete calculation of CSB effects at LO and extracted the strong contribution to the neutron-proton mass 
difference from this analysis. The value obtained, 
$\delta m_N^{\rm str} = (1.5 \pm 0.8 \ {\rm (exp.)} \pm 0.5 \ {\rm (th.))} 
\ {\rm MeV}$, 
is consistent with the result based on the Cottingham sum rule and with the recent  lattice calculations. 
At present the uncertainty in this results is dominated by the
experimental error bars -- an improvement on this side would be very
important. On the other hand, a  calculation of higher order effects is also called for to confirm
 the  theoretical uncertainty estimate.

\vspace*{-0.5cm}

\acknowledgments

\vspace*{-0.3cm}

I would like to thank E. Epelbaum,  A. Filin, J. Haidenbauer, C. Hanhart, A. Kudryavtsev, V. Lensky
and U.-G. Mei\ss ner  for fruitful and enjoyable collaboration. I thank   the organizers
for the  well-organized conference and for the invitation to give this talk.
Work supported in parts by funds provided from the Helmholtz
Association (grants VH-NG-222, VH-VI-231) and by the DFG (SFB/TR 16
and DFG-RFBR grant 436 RUS 113/991/0-1) and the EU HadronPhysics2
project.  I acknowledge the support of the Federal Agency of
Atomic Research of the Russian Federation.

\end{document}